\documentclass[runningheads,pdflatex]{llncs}
\usepackage[pdftex]{graphicx}
\usepackage{comment}

\usepackage{amssymb,amsmath}
\usepackage{cancel}

\usepackage{url}

\begin{document}


\setlength\abovecaptionskip{0pt}
\setlength\textfloatsep{10pt}

\title{Operation-based Collaborative Data Sharing\\
for Distributed Systems}
\titlerunning{Operation-based Collaborative Data Sharing}

\author{Masato Takeichi}
\authorrunning{M. Takeichi}
\institute{University of Tokyo\\
\email{takeichi@acm.org}\\
}

\maketitle


%
%


\begin{abstract}
Collaborative Data Sharing raises a fundamental issue in
distributed systems.
Several strategies have been
proposed for
making shared data consistent
between peers in such a way that the shared part
of their local data
become equal.

Most of the proposals rely on {\em state-based} semantics.
But this suffers from a lack of descriptiveness
in conflict-free features of synchronization
required for flexible network connections.
Recent applications tend to use non-permanent connection
with mobile devices or allow temporary breakaways
from the system, for example.

To settle ourselves in conflict-free data sharing,
we propose a novel scheme
{\em Operation-based
Collaborative Data Sharing} that enables
conflict-free strategies for synchronization
based on operational semantics.
\end{abstract}

\section{Introduction}
\label{sec:Introduction}

Each site, or peer of distributed systems
has its exclusive property of the contents
and the policy for data sharing.
For collaborative work between peers,
the peer expects partner peers to receive some of its
data and asks them for returning the updated data,
or it asks partner peers to provide their data
for use with its local data.

This kind of data sharing is common in
our real-world systems.
Although data sharing without updates
is simple,
collaborative data sharing with the update
propagation of shared data
poses significant problems due to concurrent updates
of different instances of the same data.
Which updates should be allowed or how the update should
be propagated to all the related peers are the typical
issues to be solved.

We have been discussing
``What should be shared'' in collaborative data
sharing,
but not so much talking
about ``How should be shared''.

Concerning the ``what'',
a seminal work on
{\em Collaborative Data Sharing}
~\cite{Ives2005,Karvounarakis2013}
brought several issues upon the specification of
data to be shared.
An approach based on
the view-updating technique
with {\em Bidirectional Transformation}
\cite{Lenses,Bohannon:08,Hidaka:10,HuMT08,HSST11}
has been proved promissing.
Among others,
the {\em Dejima} 1.0 architecture
\cite{dejima-theory:2019,SFDI-Asano2020}
and the {\em BCDS Agent}~\cite{Takeichi21JSSST}
based on Bidirectional Transformation
reveal the effectiveness of using
bidirectional transformation for peers to
control the accessibility of local data.

The basic scheme of these ideas is based
on {\em state-based} semantics. That is,
data to be shared is compared with and
moved to and from between peers.
Although this is straightforward in a sense,
there may be several problems;
the size of messages
for data exchange tends to grow, and
possible conflicts occur due to
concurrent updates.

Looking from the other side,
how to share data between peers
is very similar to
how to synchronize distributed replicas to be the same.
They are almost equivalent except
original intentions.
And our problem to be solved
is how to synchronize distributed replicas in
serverless
distributed systems.

We have various kind of
{\em Conflict-free Replicated Data Types}
(CRDTs)~\cite{Shapiro2011CRDTs}.
The CRDT approach restricts
available operations acted on replicated data;
the {\em Grow-Only-Set} (G-Set) CRDT allows only
the insertion operation on the set data, for example.
Concerning this, based on operational semantics,
we propose another schema for conflict-free
data sharing using {\em effectful}
operations that enables us to insert and delete
elements as we like~\cite{CCSS2021:Takeichi}.

In this paper, we will explore a novel scheme
for collaborative data sharing
based on {\em operational} approach.
The semantics of collaborative data sharing
is redefined using operations performed on
peer's local data.
And as a natural course, operations are exchanged
each other for making effective data sharing between
peers.

Our {\em Operation-based Collaborative Data Sharing}
(OCDS)
can solve the problem concerning possible
conflicts between concurrent operations
by conflict-free synchronization for eventual
consistency.
And this accepts more operations than
CRDTs.
It is the most remarkable feature of our OCDS
compared with CRDTs.

\section{Operations
and Transformations in Data Sharing}
\label{sec:OprationTransformation}

\begin{figure}[htb]
 \centering
  \includegraphics
  [width=0.75\linewidth]{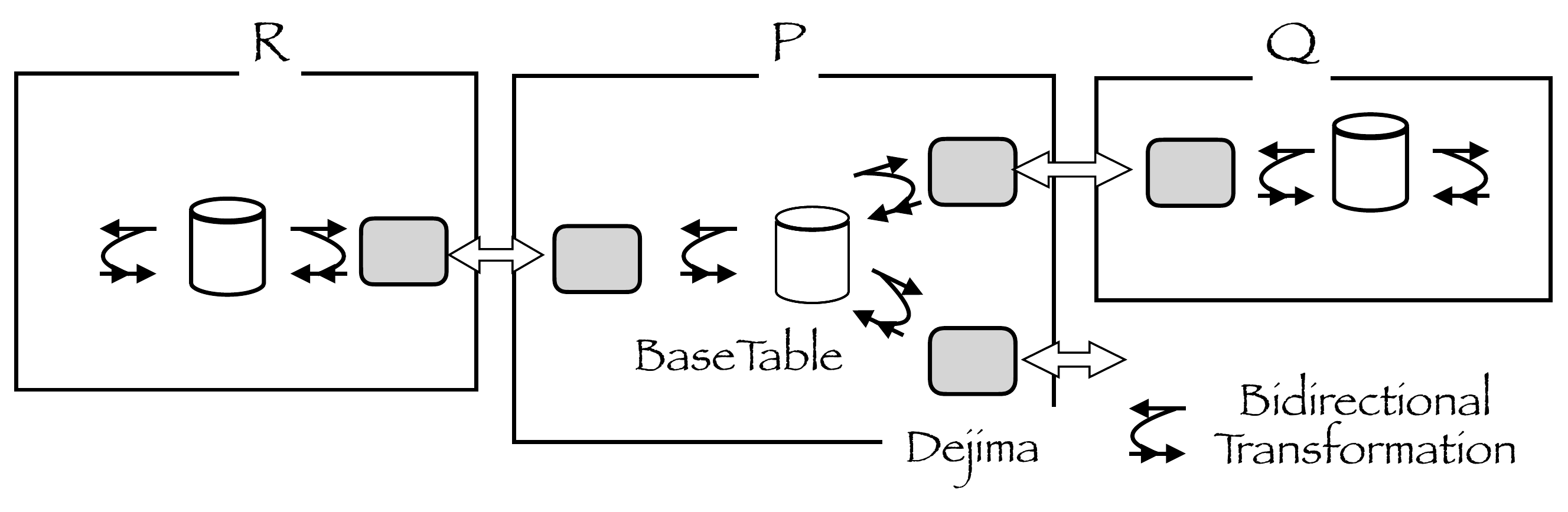}
  \vspace*{0pt}
  \caption{Dejima Architecture for
  Collaborative Data Sharing}
  \label{fig:Dejima}
\end{figure}

The {\em Dejima} architecture mentioned in the previous
section configures peers with local data called
{\em Base Table} and several additional
{\em Dejima Tables}.
The shared data is located both in the Dejima Tables in
peers $P$ and $Q$ as illustrated in
Fig.\ref{fig:Dejima}.
{\em Bidirectional Transformation} is employed to
convert data between the Base Table
and the Dejima Table. It controls what to
provide and what to accept for data sharing.

As described above, the Dejima architecture relies on
{\em state-based} semantics.

We will give another definition of {\em collaborative
data sharing} based on operations
on the local data.
We assume that the configuration of peers
is the same as the Dejima
architecture without the Dejima Table which
corresponds to
the state of shared data.

\subsubsection*{Updating Operation}
Peer $P$ has local data $D_P$ of (structured) type
$\mathcal{D}_P$ with operations on $\mathcal{D}_P$.
The operation takes postfix form
$\odot_P p::\mathcal{D}_P \rightarrow \mathcal{D}_P$
where $p \in{\cup_{D_P\in \mathcal{D_P}}}D_P$,
which maps $D_P$ to $D_P \odot_P p \in \mathcal{D}_P$,
i.e., $D_P \mapsto D_P \odot_P p$.

From operational point of view,
``$\odot_P p$ updates $D_P$ into $D_P \odot_P p$''.
Here, $p$ is not necessarily a single element, but may be composed of several elements in
${\cup_{D_P\in \mathcal{D_P}}}D_P$.
But, for simplicity, we write them as a single $p$.

Operation $\odot_P p$ is derived from operator $\odot_P$
in $P$ indexed by $p$.
Operator $\odot_P$ stands for a generic symbol for
$\oplus$, $\ominus$, $\otimes$, $\oslash$, etc.
For convenience, we use a postfix identity operation
``$!$'' which does not change $D_P$, i.e., $D_P!=D_P$
for any $D_P \in \mathcal{D}_P$.

\subsubsection*{Transformation Function}
In addition to these operations, transformation functions
$\langle get_P^q, put_P^q \rangle$,
where
$get_P^q::\mathcal{D}_P \rightarrow \mathcal{D}$ and
$put_P^q::\mathcal{D} \rightarrow \mathcal{D}_P$
for some $\mathcal{D}$.

With a reservation, $put_P^q$ may use $\mathcal{D}_P$
along with $\mathcal{D}$ as
$put_P^q:: \mathcal{D}_P \times
\mathcal{D} \rightarrow \mathcal{D}_P$.
This will be clear shortly with
the reason why $\mathcal{D}_P$
appears in the domain.

Same for the partner peer $Q$ with
$\mathcal{D}_Q$ and
$\langle get_Q^p, put_Q^p \rangle$,
and $\mathcal{D}$
appeared in definitions in $Q$ is common to $\mathcal{D}$
in $P$.
Thus, $\mathcal{D}$ ``combines'' $P$ and $Q$
as a connector for data exchange.

Suffixes for identifying the peer, e.g., $P$
in $\mathcal{D}_P$, $D_P$, $\odot_P$, ...
are omitted when they are clear from the context.

\subsubsection*{Properties of Transformation Functions}
Given data $D_P \in \mathcal{D}_P$ in $P$,
$get_P^q(D_P)$ gives some $D \in \mathcal{D}$.
And then, $Q$ gets share with data
$D_Q=put_Q^p(D)\in \mathcal{D}_Q$
which corresponds to $D_P$ in $P$.
And reciprocally, $P$ shares
$D_P'=put_P^q(D')=put_P^q(get_Q^p(D_Q'))\in \mathcal{D}_P$
which corresponds to $D_Q'$ in $Q$.

Intuitively, we may understand that
part of $D_P$ and part of $D_Q$ are shared
each other.

If it happened to be $D=D'$,
it is natural to assume that $D_P=D_P'$
and $D_Q=D_Q'$ hold with the above equalities,
so
\begin{equation*}
  (put_P^q\cdot get_Q^p)\cdot(put_Q^p\cdot get_P^q)=
  (put_Q^p\cdot get_P^q)\cdot(put_P^q\cdot get_Q^p)=
  id
\end{equation*}
hold.
Then, what should we require for
$get$ and
$put$
in each peer?

Considering that
$\langle get_P^q, put_P^q \rangle$ and
$\langle get_Q^p, put_Q^p \rangle$
are prepared independently in $P$ and $Q$,
it is reasonable to ask for
\begin{equation*}
  \begin{split}
    put_P^q \cdot get_P^q &= get_P^q \cdot put_P^q = id\\
    put_Q^p \cdot get_Q^p &= get_Q^p \cdot put_Q^p = id
  \end{split}
\end{equation*}
This is what we call the ``Round-tripping'' property of
{\em well-behaved} bidirectional transformation.
And we require our $get$ and $put$ to satisfy
this property.

To define well-behaved bidirectional transformation
$\langle get_P^q, put_P^q \rangle$,
taking $\mathcal{D}_P$ along with $\mathcal{D}$
as the domain of
$put_P^q$
is of a great help to define well-behaved transformation.
From this reason, we sometimes define it as
$put_P^q:: \mathcal{D}_P \times
\mathcal{D} \rightarrow \mathcal{D}_P$.

From our operational viewpoint, this $put_P^q$
updates the current instance $D_P$
of mutable data $\mathcal{D}_P$ with
$D\in \mathcal{D}$
to produce a new instance
$D_P' \in \mathcal{D}_P$.
This is natural and reasonable
in that we may use the current data when updating
mutable data.

\section{Operation-based Collaborative Data Sharing}
\label{sec:Op-basedCCDS}

Local operation $\odot_P p$
causes an effect on elements of structured data $D_P$
at a time, and therefore $D_P \odot_P p$
is a new instance $D_P'\in \mathcal{D}_P$
which is almost the same as $D_P$
except for some different elements.
A simple example of $\mathcal{D}_P$ is the set with
standard operations ``{\sf insert} an element $p$''
(written as $\cup\{p\}$) and
``{\sf delete} an element $p$''($\setminus\{p\}$).

As for collaborative data sharing between $P$ and $Q$,
a straightforward method for synchronization
would be to exchange $D_P$ and $D_Q$ through $D$
with transformation by $get$s and $put$s
at the gateways of $P$ and $Q$.
This approach is called ``state-based'' because
the state of the data is wholly concerned
in discussion.

Although the state-based approach to collaborative
data sharing is most common,
it is not suitable for {\em conflict-free}
strategies that aim to do something gradually
in $P$ and $Q$ for the shared part of $D_P$ and
the part of $D_Q$ to arrive at the same state
eventually.
The conflict-free approach liberates us from the
necessity of
global locks for exclusive
access to the whole distributed data to avoid
conflicts between concurrent updates.
This is particularly useful in distributed systems
with no coordination by any peers such as P2P-configured
or composed of highly independent peers.

While the {\em Conflict-free Replicated Data Type}
(CRDT) restricts operations so that the data in each
peer can be easily merged,
our conflict-free approach allows a wider class of
operations that are common to general data structures.
Recently, a novel scheme for
{\em Conflict-free Collaborative Set Sharing}
\cite{CCSS2021:Takeichi} is proposed
using operations performed so far instead of
directly merging the current data.
Although this concentrates on the set data,
it can be extended to our data sharing where
transformations lie between peers' local data.

\subsection{Homomorphic Data Structures for Data Sharing}
\label{sec:HomDataStructure}

If
$D_Q=put_Q^p(get_P^q(D_P))$
and
$D_P=put_P^q(get_Q^p(D_Q))$
hold,
we say that ``$D_P$ and $D_Q$ are {\em consistent}''
and write this as $D_P\sim D_Q$.
In other words, consistent $D_P$ and $D_Q$ have
corresponding parts which are shared each other
through intermediate data $D$ between them.

Assuming that $D_P\sim D_Q$,
then what happens when operation
$\odot_P p$ is performed on $D_P$ to
produce $D_P\odot_P p$?
\begin{itemize}
  \item If $get_P^q(D_P\odot_P p)$ gives some
  $D'$ which is to be transformed next by $put_Q^p$, and
  \begin{itemize}
    \item If $put_Q^p(D')$ gives
    some $D_Q'\in \mathcal{D}_Q$,
    then $D_P\odot_P p \sim D_Q'$.
    \item Otherwise, $D_P\odot_P p$
    has no corresponding instance in $\mathcal{D}_Q$.
  \end{itemize}
  \item  Otherwise, $D_P\odot_P p$
  has no corresponding instance in $\mathcal{D}_Q$.
\end{itemize}

Since $\odot_P p$ changes some elements of $D_P$,
we hope that $D'$ and $D_Q'$
also change some element as $D'=D \odot x$
and $D_Q'=D_Q \odot_Q q$ with $\odot x$ and $\odot_Q q$.

In most of our data sharing applications,
$D_P$, $D_Q$ and intermediate data $D$
are {\em homomorphic} each other in that the above
conditions are satisfied.

In this respect, our transformation functions $get$
and $put$
partly provide {\em homomorphism}.
The simplest example would be the case where all the
related data structures are sets or SQL tables, etc.
In general, these are not necessarily the same but are
homomorphic.
And we need more about homomorphism on operations
for our operation-based data sharing.

\subsubsection*{Homomorphic Data Structures with
Operations}

Data type
$\langle\mathcal{A},\circledcirc_A\rangle$
is closed with respect to operations $\circledcirc_A a$
for any $a\in \cup_{A\in\mathcal{A}}A$,
where operator symbol
$\circledcirc_A :: (\mathcal{A},\cup_{A\in\mathcal{A}}A)
\rightarrow \mathcal{A}$
represents any operators in $\mathcal{A}$.
We simply write here $\circledcirc_A$ for the set of
operators in $\mathcal{A}$ and use the same symbol for
one of them as a generic operator in an overloaded manner.

The operation
$\circledcirc_A a :: \mathcal{A} \rightarrow \mathcal{A}$
is postfixed to the operand
$A \in \mathcal{A}$ to produces
$A'=A \circledcirc_A a \in \mathcal{A}$.

This models the {\em mutable} state data $A$
with operations $\circledcirc_A a$ on $A$ using some
element $a$.

\subsubsection*{Definition of Homomorphic Data Types}
Data types
$\langle\mathcal{A},\circledcirc_A\rangle$
and
$\langle\mathcal{B},\circledcirc_B\rangle$
are {\em homomorphic} if there exist
$h::\mathcal{A} \rightarrow \mathcal{B}$
and overloaded
$h::\circledcirc_A \rightarrow \circledcirc_B$
satisfying
\begin{equation*}
  \begin{split}
  &\forall A\in \mathcal{A}.
  \exists B \in \mathcal{B}. B=h (A)\\
  &\forall A\in \mathcal{A}. \forall a\in
  \cup_{A\in\mathcal{A}}A.
  \exists B \in \mathcal{B}.\exists b \in
  \cup_{B\in\mathcal{B}}.
  B \circledcirc_B b =
  h(A \circledcirc_A a )
  \end{split}
\end{equation*}

We assume that every data type
$\langle\mathcal{A},\circledcirc_A\rangle$
has an identity operation ``$!$'' which does
not affect the state of data. That is, for any
$A\in\mathcal{A}$, $A!=A$ holds.

In short,
for operation-based collaboration, we require
operations to exchange between
homomorphic data types so that operation on a peer
corresponds to operation on
the partner peer.

\subsubsection*{Examples of Homomorphic Data Types}
Previous works on state-based data sharing
with transformation
\cite{dejima-theory:2019,SFDI-Asano2020,Takeichi21JSSST}
exclusively deal with SQL databases as local data.
Specifically,
if the intermediate data $D$ is defined
as the view of the SQL
table $D_P$ of the local data,
it is obvious that $D_P$ and $D$ are homomorphic
because $D$ is produced by selection and projection of
$D_P$.
The Dejima architecture allows so-called SPJU
(Select-Project-Join-Union) queries by the SQL's
{\sf SELECT-FROM-WHERE-UNION} construct
for the view. But, since it is not clear what
operations
are permitted on (multiple) SQL tables of $D_P$,
we need more to work on
making sure that $D_P$ and $D$ are
homomorphic. We leave this for the future.

As a demonstration of the independence of implementation
of the local data $D_P$ from the intermediate data $D$
of our operation-based data sharing,
consider the case that $D$ is a set, i.e.,
no duplicates in aggregation, and $D_P$ implements
set by the binary search tree.
In this case, we easily give a homomorphism mapping
from $D_P$ to $D$.
Or, it should be grounded in the data abstraction
mechanism.

As for the relationship of homomorphic data types
with state machines, see the Appendix.

\subsection{Transformation of Operations}
\label{sec:TransOp}

For homomorphic data structures
$\langle\mathcal{D}_P,\odot_P\rangle$,
$\langle\mathcal{D},\odot\rangle$
and
$\langle\mathcal{D}_Q,\odot_Q\rangle$,
operations are transformed according to
the hompmorphisms by $get$ and $put$.
We write
\begin{verse}
  $\odot_P {}_P^q\!{\rightarrowtail} \odot x$,
  if $get_P^q(D_P\odot_P p)$ gives $D \odot x$.\\
  $\odot x ~{\looparrowright}_Q^p\odot_Q q$,
  if $put_Q^p(D\odot x)$ gives $D_Q \odot_Q q$
\end{verse}

\begin{figure}[htb]
 \centering
  \includegraphics
  [width=0.75\linewidth]{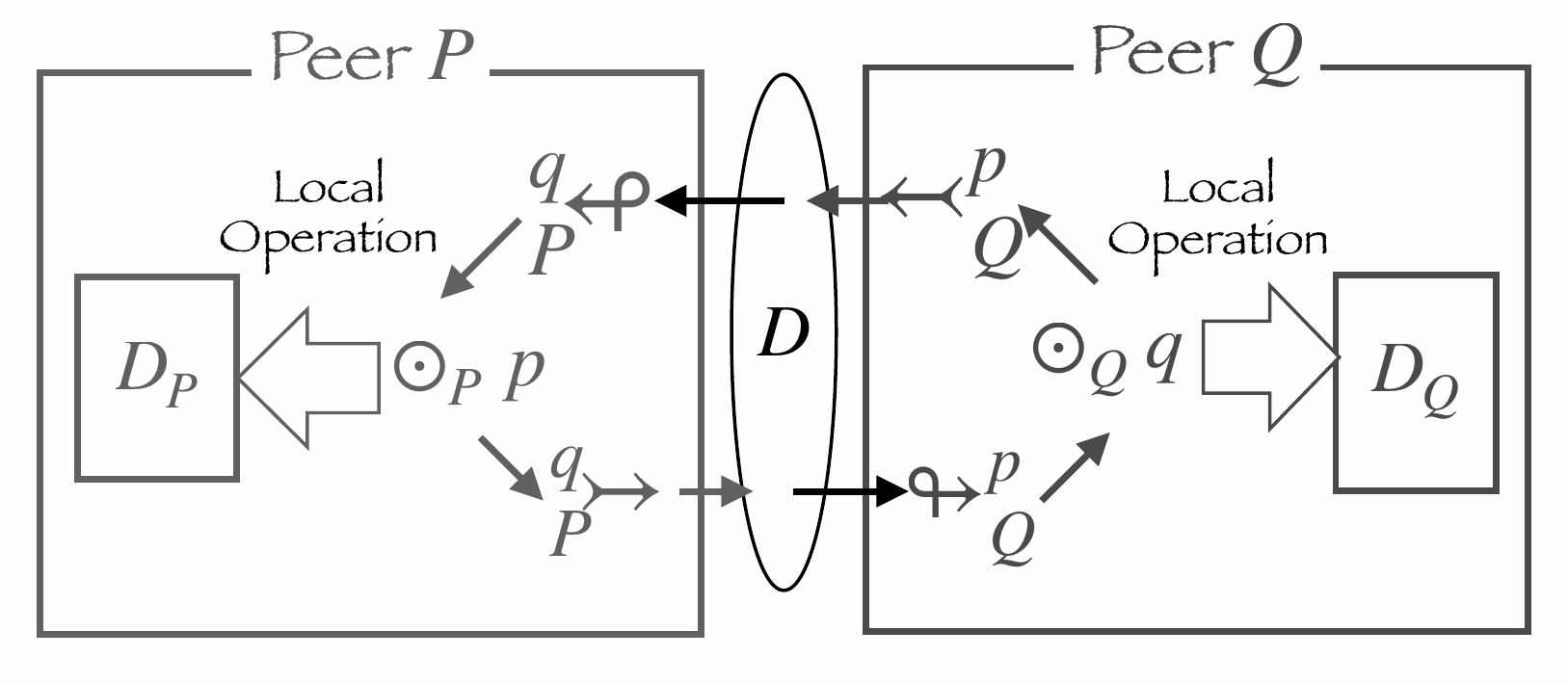}
  \vspace*{0pt}
  \caption{Operation-based Collaborative Data Sharing}
  \label{fig:Op-basedCDS}
\end{figure}

Our Operation-based Collaborative Data Sharing wholly
sends and receives operations instead of data as shown
in the diagram of Fig.\ref{fig:Op-basedCDS}.

In this way, peers
communicate
updating operations to and from each other with
necessary transformation at the gateway of the peer.


\section{Architecture for
Collaborative Data Sharing}
\label{sec:Architecture}

Peers of our Collaborative Data Sharing
run concurrently and they transmit their updates
asynchronously to and from each other.
Thus, the {\em payload} of the communication message
between peers is the updating operation
from a peer to its partner peer.

The peer as the {\em client} sends local operations
to the partner peers,
and as the {\em server} receives remote operations
from the partners and then perform necessary operations
to reflect them on the local data.

In these processes, each peer works as follows.
Peer $P$ asynchronously receives local operations
from the user and remote operations from
the partner peers.
These operations are to be performed on $D_P$
and
are stored in the queue for serialized access to $D_P$.

So far, we explained our scheme
solely with peers $P$ and $Q$.
But, in general, each peer $P$ has
multiple peers connected
in the system.
For $P$ to do with all the partmer peers,
we need to clarify how $P$ should do.

In peer $P$, every update on data $D_P$
is propagated to all the partner peers
$k=\cdots, Q, \cdots$
through the outgoing communication ports prepared
for each peer $k$ after it is transformed by
${}_P^k\!{\rightarrowtail}$.
And remote operations from peers are received
asynchronously from the partner peers
$K=\cdots, Q, \cdots$ through the incoming ports each
prepared for the peer $k$ and transformed by
${\looparrowright}_P^k$.

As the local and remote operations arrive asynchronously,
the peer needs to provide queues for them to perform
the operations on
the local data.

In addition to this,
we follow the scheme of {\em Conflict-free Collaborative
Set Sharing} described in \cite{CCSS2021:Takeichi}
for conflict-free synchronization.


We call
the implementation of the peer as described above
by the name ``OCDS Agent''.

\begin{figure}[htb]
 \centering
  \includegraphics
  [width=0.75\linewidth]{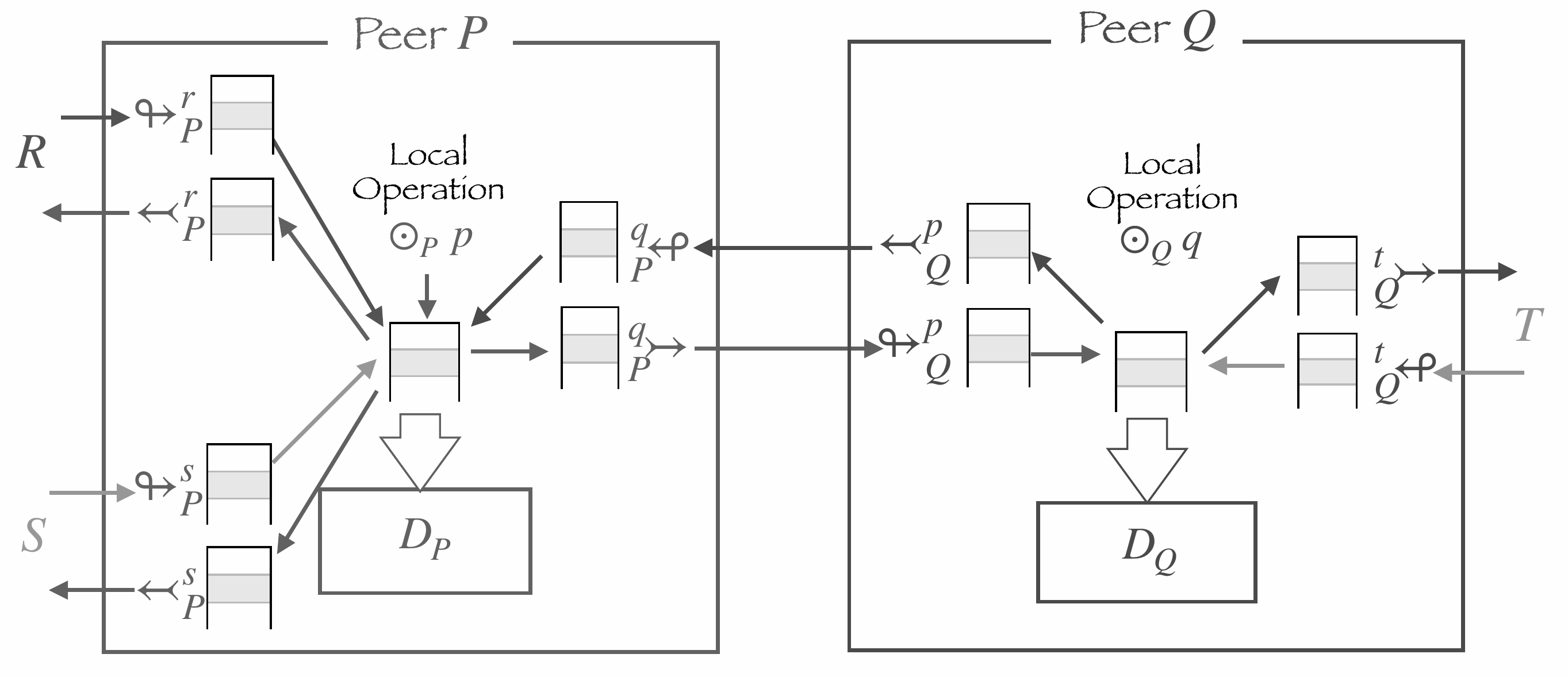}
  \vspace*{0pt}
  \caption{OCDS Agent for
  Operation-based Collaborative Data Sharing}
  \label{fig:OCDSAgent}
\end{figure}

The OCDS Agent is developed to
achieve conflict-free synchronization
of the local data using internal queues
for serialization of asynchronous access
to the local data and asynchronous
transmission of operations as
illustrated in Fig.\ref{fig:OCDSAgent}.

\section{An Example of Operation-based
Collaborative Data Sharing}
\label{sec:Example}

\subsubsection*{Sharing Double and Triple Numbers}
Let $D_P$ be a set of integers with operations
``{\sf insert} an element $p$''
($\cup\{p\}$) and
``{\sf delete} an element $p$''($\setminus\{p\}$).

Bidirectional transformation defied in $P$ is
\begin{equation*}
  \begin{split}
    get_P^p(D_P)&=\{p~|~p\%2=0, ~p \in D_P\}\\
    put_P^q(D_P,D)&=D_P \setminus
    get_P^q(D_P) \cup \{x~|~x\%2=0, ~x \in D\}
  \end{split}
\end{equation*}
where $\%$ represents the modulo operation.

Functions $get_P^q$ and $put_P^q$ define the mapping for
view-updating of the state-based approach;
$get_P^q$ produces the view $D=get_P^q(D_P)$, and
$put_P^q$ replects the update $D'$ of $D$ onto the
source as $D_P'=put_P^q(D_P,D)$.
We can see that this bidirectional transformation
$\langle get_P^q, put_P^q \rangle$
satisfies the round-tripping property and so
is well-behaved.

Similarly, defined in $Q$:
\begin{equation*}
  \begin{split}
    get_Q^p(D_Q)&=\{q~|~q\%3=0, ~q \in D_Q\}\\
    put_Q^p(D_Q,D)&=D_P \setminus
    get_Q^p(D_Q) \cup \{x~|~x\%3=0, ~x \in D\}.
  \end{split}
\end{equation*}

Then, we use these for data sharing in a way that the
intermediate data $D$ represents shared data
consisting elements in both of $get_P^q(D_P)$
and $get_Q^p(D_Q)$, i.e.,
$D=get_P^q(D_P)\cap get_Q^p(D_Q)$.
In brief, $D$ contains sextuple numbers,
i.e., numbers divisible by 6, common to $D_P$ and $D_Q$.

We can confirm by the state-based semantics
that local updates in $P$ and $Q$ are faithfully
reflected in both $D_P$ and $D_Q$ through
the Dejima $D$ if this condition holds.


Now, we are going to our operation-based sharing.

Recall that $get_P^q$ tells us that
$P$ is willing to share double numbers with $Q$,
and that $get_Q^p$ tells us that $Q$ is willing
to share triple numbers with $P$.
However, the $put$ functions tell us that
$P$ will accept only double numbers, and
$Q$ will accept only triple numbers from the
common intermediate data $D$.

A short story is here:
\begin{enumerate}
  \item Start from $D_P=\{1,2,3,4\}$
  and $D_Q=\{2,3,4,9\}$.
  They are consistent, i.e., $D_P\sim D_Q$ since $D=\{\}$.
  \item Network connection fails.
  \item Concurrently, $P$ does $\cup\{6\}$
  and $Q$ does $\setminus \{4\}$.
  \item Connection restored, and synchronization
  processes start
  in $P$ and $Q$ independently.
\end{enumerate}

Then, what happens in synchronization processes?

These operations are in fact {\em effectful}\footnote{
The concept of the ``effectful'' operation
is described in \cite{CCSS2021:Takeichi}}
in that $\cup \{6\}$ is applied to $D_P$
which does not contain $6$,
and $\setminus \{4\}$ is applied to $D_Q$ which
does contain $4$.

In Step 3, $P$'s local data becomes
$D_P'=D_P\cup\{6\}=\{1,2,3,4,6\}$,
and $Q$'s local data becomes
$D_Q'=D_Q\setminus \{4\}=\{2,3,9\}$.

Synchronization proceeds as
\begin{itemize}
  \item Since
  $\cup \{6\}{}_P^q\!{\rightarrowtail}\cup
   \{6\}{\looparrowright}_Q^p$, $6$ is added to
   $D_Q'$ to produce
   $D_Q''=D_Q'\cup \{6\}= \{2,3,6,9\}$.
  \item On the other direction,
  since $\setminus\{4\}$ in $Q$ cannot be passed
  to ${}_Q^p\!{\rightarrowtail}$ because
  $get_Q^p$ rejects $4$,
  this operation does not arrive at $P$.
\end{itemize}

Thus, these synchronization processes concurrently
done in $P$ and $Q$ lead $P$'s data and $Q$'s data
to the consistent state,
i.e., $D_P'\sim D_Q''$ with $D=\{6\}$.


Another story is here:
In Step 3 above, what happens if
``$Q$ does $\setminus \{6\}$''
instead of
``$Q$ does $\setminus \{4\}$''?

Note that these operations are not effectful because
$\setminus \{6\}$ here
is applied to $D_Q$ which does not contain $6$.
During the period of network failure,
$P$’s local data becomes
$D_P'=D_P\cup\{6\}=\{1,2,3,4,6\}$ as before,
and $Q$’s local data remains at it has been because
$D_Q'=D_Q\setminus\{6\}=\{2,3,4,9\}$.

Synchronization proceeds as follows
after the network connection
is restored.
\begin{itemize}
  \item $\cup \{6\}{}_P^q\!{\rightarrowtail}\cup
   \{6\}{\looparrowright}_Q^p$
   causes changes
   $D_Q''=D_Q'\cup \{6\}= \{2,3,6,9\}$
   as the previous case.
  \item And since
  $\setminus \{6\}{}_Q^p\!{\rightarrowtail}
  \setminus \{6\}{\looparrowright}_P^q\setminus \{6\}$,
  $P$ may produces a new state
  $D_P''=D_P'\setminus\{6\}=\{1,2,3,4\}$.
\end{itemize}
If the synchronization in $P$ proceeds as above,
$P$ loses $6$ which was added in Step 3,
while it is added to $Q$'s local data by $Q$'s
synchronization.
This breaks the consistency of
$D_P''$ and $D_Q'$.

From these examples,
we observe that the effectful set operations
in concurrent updates is essential
for conflict-free synchronization.
They effectively avoid insertion/deletion
conflicts in synchronization.
This is an extension of the scheme
for data sharing described in \cite{CCSS2021:Takeichi}.
Here, we used transformations
$get$ and $put$ at the gateways of the peers.

\section{Remarks}
\label{sec:Remarks}
We can employ our OCDS Agent for configuring
serverless distributed systems with
ensuring eventual consistency of peers' local data.

We may consider this as an alternative scheme for
the Dejima style data sharing~\cite{dejima-theory:2019}
that has been
implemented to ensure the global strong
consistency by locking on the way during the update
propagation.
Our conflict-free approach allows peers to
leave and join at any time and can afford to
the network failure and restoration.

A remarkable feature of our OCDS
is that it enables us to control the data;
what to provide and
what to accept for sharing with other peers.
This contrasts clearly with other conflict-free
data sharing or data synchronization of replicated
data such as CRDTs.

\bibliographystyle{abbrv}
\bibliography{citation}

\newpage
\noindent
{\bf Appendix}

\subsubsection*{Example of Homomorphic States}

\begin{figure}[htb]
 \centering
  \includegraphics
  [width=0.75\linewidth]{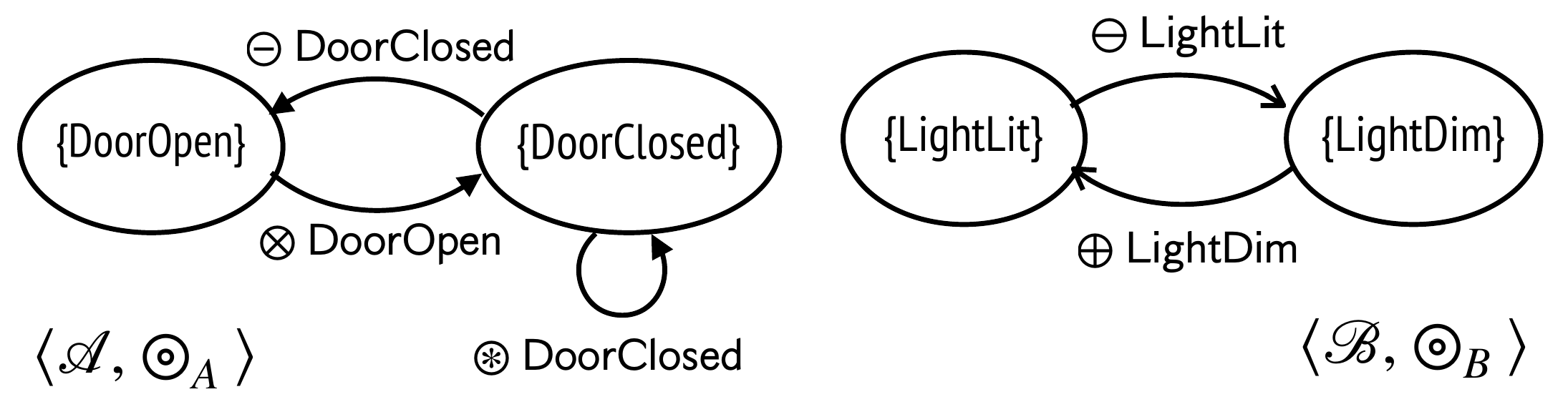}
  \vspace*{0pt}
  \caption{Collaborative Working Door and Light}
  \label{fig:DoorLight}
\end{figure}

Let
$\langle\mathcal{A},\circledcirc_A\rangle$
show the state of the door at the entrance and
$\langle\mathcal{B},\circledcirc_B\rangle$
show the state of the electric
light of the entrance hall.
\begin{verse}
  $\mathcal{A}=\{\{{\sf DoorOpen}\},
  \{{\sf DoorClosed}\}\}$\\
  $\circledcirc_A:$ $\circleddash$ for Open,
  $\otimes$ for Close,
  $\circledast$ for RingBell.
\end{verse}
\begin{verse}
  $\mathcal{B}=\{\{{\sf LightLit}\},
  \{{\sf LightDim}\}\}$\\
  $\circledcirc_B:$ $\oplus$ for On,
  $\ominus$ for Off.
\end{verse}
See the transitions in Fig.\ref{fig:DoorLight}.
Note that every operator
$\circledcirc$ takes an operand after it to
validate application, e.g., RingBell operation
$\circledast$ is valid only if the current state is
$\sf DoorClosed$.

We can define the homomorphic mapping
$h$ from
$\langle\mathcal{A},\circledcirc_A\rangle$
to
$\langle\mathcal{B},\circledcirc_B\rangle$.
\begin{itemize}
  \item For $\mathcal{A}$ to $\mathcal{B}$,
  \begin{equation*}
    h(\{{\sf DoorOpen}\})=\{{\sf LightLit}\},
    h(\{{\sf DoorClosed}\})=\{{\sf LightDim}\}
  \end{equation*}
  \item For $\circledcirc_A$ to $\circledcirc_B$,
  \begin{equation*}
    \begin{split}
    &h(\circleddash \{{\sf DoorClosed}\})=
    \ominus \{{\sf LightLit}\},
    h(\otimes \{{\sf DoorOpen}\})=
    \oplus \{{\sf LightDim}\},\\
    &h(\circledast \{{\sf DoorClosed}\})=!
  \end{split}
  \end{equation*}
\end{itemize}

Thus, the door and the light work together.
Along with the homomorphism for the reverse direction
gives us the collaborative updates of the states of
the door and the light.

\end{document}